\documentclass[twocolumn,aps,prc,superscriptaddress,showpacs,floatfix]{revtex4}
\usepackage{amssymb}
\usepackage{amsmath}
\usepackage{graphicx}

\begin{document}

\title{High-energy behavior of the nuclear symmetry potential in asymmetric
nuclear matter}
\author{Lie-Wen Chen}
\affiliation{Institute of Theoretical Physics, Shanghai Jiao Tong University, Shanghai
200240, China}
\affiliation{Center of Theoretical Nuclear Physics, National Laboratory of Heavy Ion
Accelerator, Lanzhou 730000, China}
\author{Che Ming Ko}
\affiliation{Cyclotron Institute and Physics Department, Texas A\&M University, College
Station, Texas 77843-3366}
\author{Bao-An Li}
\affiliation{Department of Chemistry and Physics, P.O. Box 419, Arkansas State
University, State University, Arkansas 72467-0419}
\date{\today }

\begin{abstract}
Using the relativistic impulse approximation with empirical NN scattering
amplitude and the nuclear scalar and vector densities from the relativistic
mean-field theory, we evaluate the Dirac optical potential for neutrons and
protons in asymmetric nuclear matter. From the resulting Schr\"{o}%
dinger-equivalent potential, the high energy behavior of the nuclear
symmetry potential is studied. We find that the symmetry potential at fixed
baryon density is essentially constant once the nucleon kinetic energy is
greater than about $500$ MeV. Moreover, for such high energy nucleon, the
symmetry potential is slightly negative below a baryon density of about $%
\rho =0.22$ fm$^{-3}$ and then increases almost linearly to positive values
at high densities. Our results thus provide an important constraint on the
energy and density dependence of nuclear symmetry potential in asymmetric
nuclear matter.
\end{abstract}

\maketitle

\section{introduction}

Recently, there is a renewed interest in the isovector part of the nucleon
mean-field potential, i.e., the nuclear symmetry potential, in isospin
asymmetric nuclear matter\cite%
{91bomb,97ulrych,das03,li04a,li04b,chen04,rizzo04,fuchs04,mazy04,chen05,05behera,baran05,samma05,zuo05,fuchs05,rizzo05}%
. Knowledge on the symmetry potential is important for understanding not
only the structure of radioactive nuclei and the reaction dynamics induced
by rare isotopes, but also many critical issues in astrophysics. Besides
depending on the nuclear density, the symmetry potential also depends on the
momentum or energy of a nucleon. Various microscopic and phenomenological
models, such as the relativistic Dirac-Bruechner-Hartree-Fock (DBHF) \cite%
{97ulrych,fuchs04,mazy04,samma05,fuchs05} and the non-relativistic
Bruechner-Hartree-Fock (BHF) \cite{91bomb,zuo05} approach, the relativistic
mean-field theory based on nucleon-meson interactions \cite{baran05}, and
the non-relativistic mean-field theory based on Skyrme-like interactions 
\cite{das03,05behera}, have been used to study the symmetry potential.
However, the results predicted by these models vary widely. In particular,
while most models predict a decreasing symmetry potential with increasing
nucleon momentum albeit at different rates, a few nuclear effective
interactions used in some of the models lead to the opposite conclusion.
Thus, any constraint on the momentum and density dependence of the symmetry
potential is very useful.

In the optical model based on the Dirac phenomenology, elastic
nucleon-nucleus scattering is described by the Dirac equation for the motion
of a nucleon in a relativistic potential. For spherical nuclei, good
agreements with experimental data were obtained in the relativistic approach
with a scalar potential (nucleon scalar self-energy) and the zeroth
component of a vector potential (nucleon vector self-energy), while the
standard non-relativistic optical model using the Schr\"{o}dinger equation
failed to describe simultaneously all experimental observables \cite%
{arnold79}. Motivated by the success of the Dirac phenomenology, a
microscopic relativistic model based on the impulse approximation (RIA) \cite%
{mcneil83,shepard83,clark83,miller83} was developed, and it was able to fit
very well the data from p+$^{40}$Ca and p+$^{208}$Pb elastic scattering at
nucleon energies of both $500$ and $800$ MeV. A nice feature of the RIA is
that it permits very little phenomenological freedom in deriving the Dirac
optical potential in nuclear matter. The basic ingredients in this method
are the free invariant nucleon-nucleon (NN) scattering amplitude and the
nuclear scalar and vector densities in nuclear matter. This is in contrast
to the relativistic DBHF approach, where different approximation schemes and
methods have been introduced for determining the Lorentz and isovector
structure of the nucleon self-energy \cite%
{97ulrych,fuchs04,mazy04,samma05,fuchs05}.

In the present work, we evaluate the Dirac optical potential for neutrons
and protons in asymmetric nuclear matter based on the relativistic impulse
approximation by using the NN scattering amplitude determined by McNeil,
Ray, and Wallace \cite{mcneil83prc}, which has been shown to be valid for
nucleons with kinetic energy greater than about $300$ MeV. The high energy
behavior of the nuclear symmetry potential from the resulting Schr\"{o}%
dinger-equivalent potential is then investigated without adjustable
parameters. We find that the nuclear symmetry potential at fixed density
becomes almost constant for nucleon kinetic energy greater than about $500$
MeV. For such high energy nucleons, the nuclear symmetry potential is
further found to be weakly attractive below a nuclear density of about $\rho
=0.22$ fm$^{-3}$ but to become increasingly repulsive when the nuclear
density increases.

The paper is organized as follows. In Section \ref{optical}, we briefly
review the relativistic impulse approximation for nuclear optical potential
and the relativistic mean-field model for nuclear scalar and vector
densities. Results on the relativistic nuclear optical potential and the
nuclear symmetry potential in asymmetric nuclear matter are presented in
Section \ref{results}. A short summary is then given in Section \ref{summary}%
.

\section{nuclear optical potential}

\label{optical}

\subsection{Relativistic impulse approximation}

Many theoretical studies have suggested that nucleon-nucleus scattering at
sufficient high energy can be viewed as the projectile nucleon being
scattered from each of the nucleons in the target nucleus. One thus can
describe the process by using the NN scattering amplitude and the ground
state nuclear density distribution of the target nucleus. For the
Lorentz-invariant NN scattering amplitude, it can be written as 
\begin{equation}
\widehat{F}=F_{S}+F_{V}\gamma _{1}^{\mu }\gamma _{2\mu }+F_{T}\sigma
_{1}^{\mu \nu }\sigma _{2\mu \nu }+F_{P}\gamma _{1}^{5}\gamma
_{2}^{5}+F_{A}\gamma _{1}^{5}\gamma _{1}^{\mu }\gamma _{2}^{5}\gamma _{2\mu }
\end{equation}%
in terms of the scalar $F_{S}$, vector $F_{V}$, tensor $F_{T}$, pseudoscalar 
$F_{P}$, and axial vector $F_{A}$ amplitudes. In the above, subscripts $1$
and $2$ distinguish Dirac operators in the spinor space of the two
scattering nucleons and $\gamma ^{\prime }$s are gamma matrices. The five
complex amplitudes $F_{S}$, $F_{V}$, $F_{T}$, $F_{P}$, and $F_{A}$ depend on
the squared momentum transfer $\mathbf{q}^{2}$ and the invariant energy of
the scattering nucleon pair, and they were determined in Ref. \cite%
{mcneil83prc} directly from the NN phase shifts that were used to
parameterize the NN scattering data. For a spin-saturated nucleus, only the
scalar ($F_{S}$) and the zeroth component of the vector ($F_{V}\gamma
_{1}^{0}\gamma _{2}^{0}$) amplitudes dominate the contribution to the
optical potential. In the relativistic impulse approximation, the optical
potential in momentum space is thus obtained by multiplying each of these
two amplitudes with corresponding momentum-space nuclear scalar ${\tilde{\rho%
}}_{S}(\mathbf{q})$ and vector $\tilde{\rho}_{V}(\mathbf{q})$ densities,
i.e., 
\begin{equation}
{\tilde{U}}_{\text{opt}}(\mathbf{q})=\frac{-4\pi ip_{\text{lab}}}{M}[F_{S}(q)%
{\tilde{\rho}}_{S}(\mathbf{q})+\gamma _{0}F_{V}(q)\tilde{\rho}_{V}(\mathbf{q}%
)],  \label{trhom}
\end{equation}%
where $p_{\text{lab}}$ and $M$ are, respectively, the laboratory momentum
and mass of the incident nucleon. The optical potential in coordinator space
is then given by the Fourier transformation of ${\tilde{U}}_{\text{opt}}(%
\mathbf{q})$, similar to the \textquotedblleft $t\rho $\textquotedblright\
approximation used in non-relativistic impulse approximation \cite%
{mcneil83prc}.

The $\mathbf{q}$-dependence of the relativistic NN amplitude is important
for calculating observables of a nucleon scattering off finite nuclei within
the Dirac phenomenology. In the present work, we are interested in the
strength of the Dirac optical potential of nucleons in infinite nuclear
matter. Since the scalar and vector densities in coordinate space are
constant in infinite nuclear matter, they are delta functions in momentum
space, i.e., $\sim \delta ^{(3)}(\mathbf{q})$. In this case, only the
forward NN scattering amplitudes, i.e., $F_{S0}\equiv F_{S}(q=0)$ and $%
F_{V0}\equiv F_{V}(q=0)$, contribute to the Fourier transform of Eq. (\ref%
{trhom}), and the nuclear coordinate-space optical potential takes the
simple form \cite{mcneil83}: 
\begin{equation}
U_{\text{opt}}=\frac{-4\pi ip_{\text{lab}}}{M}[F_{S0}\rho _{S}+\gamma
_{0}F_{V0}\rho _{V}],  \label{trhor}
\end{equation}%
where $\rho _{S}$ and $\rho _{V}$ are, respectively, the spatial scalar and
vector densities of an infinite nuclear matter.

The optical potential $U_{\text{opt}}$ is a $4\times 4$ matrix in the Dirac
spinor space of the projectile nucleon and includes following scalar $U_{S}^{%
\mathrm{tot}}$ and vector $U_{0}^{\mathrm{tot}}$ pieces: 
\begin{equation}
U_{\text{opt}}=U_{S}^{\mathrm{tot}}+\gamma _{0}U_{0}^{\mathrm{tot}}.
\end{equation}%
Since $U_{S}^{\mathrm{tot}}$ and $U_{0}^{\mathrm{tot}}$ are generally
complex, they can be expressed in terms of their real and imaginary parts,
i.e., 
\begin{equation}
U_{S}^{\mathrm{tot}}=U_{S}+iW_{S},\text{ \ }U_{0}^{\mathrm{tot}%
}=U_{0}+iW_{0},
\end{equation}

\subsection{Nuclear scalar densities}

To evaluate the RIA optical potential for neutrons and protons, we need the
values for the isospin-dependent $F_{S0}$ and $F_{V0}$, which can be found
in Ref. \cite{mcneil83prc}, as well as the scalar and vector densities for
neutrons and protons. For the latter, we determine them using the
relativistic mean-field (RMF) theory \cite{rmf} with a Lagrangian density
that includes the nucleon field $\psi $, the isoscalar-scalar meson field $%
\sigma $, the isoscalar-vector meson field $\omega $, the isovector-vector
meson field $\rho $, and the isovector-scalar meson field $\delta $, i.e., 
\begin{eqnarray}
&&{\mathcal{L}}(\psi ,\sigma ,\bm{\omega},\bm{\rho},\delta )=\bar{\psi}\left[
\bm{\gamma}_{\mu }(i\partial ^{\mu }-g_{\omega }\bm{\omega}^{\mu
})-(M-g_{\sigma }\sigma )\right] \psi   \notag \\
&&+\frac{1}{2}(\partial _{\mu }\sigma \partial ^{\mu }\sigma -m_{\sigma
}^{2}\sigma ^{2})-\frac{1}{4}\bm{\omega}_{\mu \nu }\bm{\omega}^{\mu \nu }+%
\frac{1}{2}m_{\omega }^{2}\bm{\omega}_{\mu }\bm{\omega}^{\mu }  \notag \\
&&-\frac{1}{3}b_{\sigma }M{(g_{\sigma }\sigma )}^{3}-\frac{1}{4}c_{\sigma }{%
(g_{\sigma }\sigma )}^{4}+\frac{1}{4}c_{\omega }{(g_{\omega }^{2}\bm{\omega}%
_{\mu }\bm{\omega}^{\mu })}^{2}  \notag \\
&&+\frac{1}{2}(\partial _{\mu }\delta \partial ^{\mu }\delta -m_{\delta }^{2}%
{\delta }^{2})+\frac{1}{2}m_{\rho }^{2}\bm{\rho}_{\mu }.\bm{\rho}^{\mu }-%
\frac{1}{4}\bm{\rho}_{\mu \nu }.\bm{\rho}^{\mu \nu }  \notag \\
&&+\frac{1}{2}(g_{\rho }^{2}\bm{\rho}_{\mu }.\bm{\rho}^{\mu })(\Lambda
_{S}g_{\sigma }^{2}\sigma ^{2}+\Lambda _{V}g_{\omega }^{2}\bm{\omega}_{\mu }%
\bm{\omega}^{\mu })  \notag \\
&&-g_{\rho }\bm{\rho}_{\mu }\bar{\psi}\gamma ^{\mu }\bm{\tau}\psi +g_{\delta
}\delta \bar{\psi}\bm{\tau}\psi \;,  \label{lag}
\end{eqnarray}%
where the antisymmetric field tensors $\omega _{\mu \nu }$ and $\rho _{\mu
\nu }$ are given by $\bm{\omega}_{\mu \nu }\equiv \partial _{\nu }\bm{\omega}%
_{\mu }-\partial _{\mu }\bm{\omega}_{\nu }$ and $\text{ }\bm{\rho}_{\mu \nu
}\equiv \partial _{\nu }\bm{\rho}_{\mu }-\partial _{\mu }\bm{\rho}_{\nu }$,
respectively, and the symbols used in Eq.(\ref{lag}) have their usual
meanings. The above Lagrangian density is general and allows us to use many
presently popular parameter sets. In the present work, we use three typical
parameter sets, namely, the very successful NL3 model \cite{lala97}, the
Z271v model, which was used to study the neutron skin of heavy nuclei and
the properties of neutron stars \cite{horowitz01}, and the HA model which
includes the isovector-scalar meson field $\delta $ and fits successfully
some results calculated with more microscopic DBHF approach \cite{bunta03}.

\begin{figure}[th]
\includegraphics[scale=0.9]{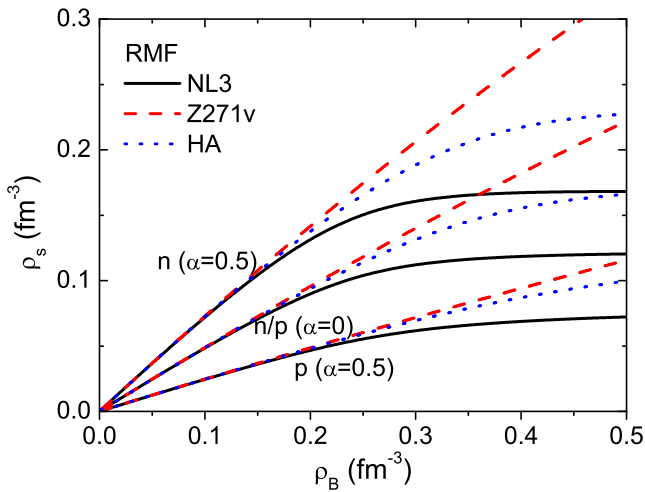}
\caption{{\protect\small (Color online) Neutron and proton scalar densities
as functions of baryon density in nuclear matter with isospin asymmetry }$%
\protect\alpha =0${\protect\small \ and }$0.5${\protect\small \ for the
parameter sets NL3, Z271v, and HA.}}
\label{rhos}
\end{figure}

In Fig. \ref{rhos}, we show the neutron and proton scalar densities $\rho _{%
\text{S}}$ as functions of the baryon density $\rho _{\text{B}}$\ (vector
density in the static infinite nuclear matter) in nuclear matter with
isospin asymmetry $\alpha =0$ and $0.5$ for parameter sets NL3, Z271v, and
HA. The isospin asymmetry is defined as $\alpha =(\rho _{n}-\rho _{p})/\rho
_{\text{B}}$ with $\rho _{\text{B}}=\rho _{n}+\rho _{p}$ and $\rho _{n}$ and 
$\rho _{p}$ denoting the neutron and proton densities, respectively. It is
seen that the neutron scalar density is larger than that of the proton at a
fixed baryon density in neutron-rich nuclear matter. While results for
different parameter sets are almost the same at lower baryon densities, they
become different when $\rho _{\text{B}}\gtrsim 0.25$ fm$^{-3}$ with Z271v
giving a larger and NL3 a smaller $\rho _{\text{S}}$ than that from the
parameter set HA. The real and imaginary parts of the scalar potential at
higher baryon densities thus depend on the interactions used in evaluating
the nucleon scalar density and have, therefore, large uncertainties.

\section{results}

\label{results}

\subsection{Relativistic nuclear optical potential}

\begin{figure}[th]
\includegraphics[scale=0.85]{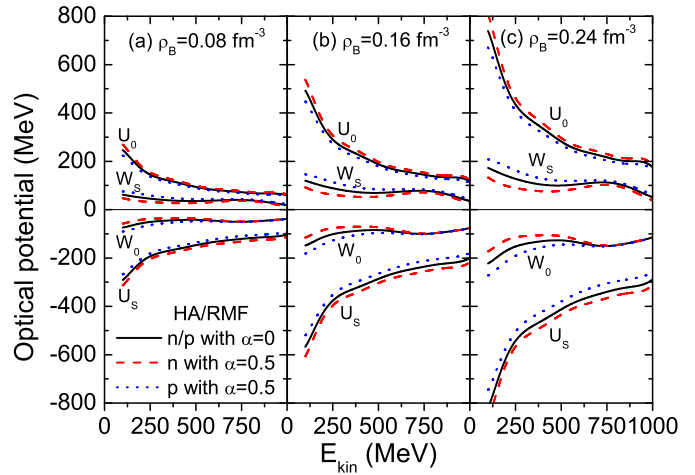}
\caption{{\protect\small (Color online) Energy dependence of real and
imaginary parts of the scalar and vector potentials for neutrons and protons
in nuclear matter with isospin asymmetry }$\protect\alpha =0${\protect\small %
\ and }$0.5${\protect\small \ for the parameter set HA.}}
\label{OPReImEkin}
\end{figure}

With neutron and proton scalar densities obtained from the RMF theory for
the parameter set HA, we have studied both the energy and density dependence
of the real and imaginary parts of the scalar and vector potentials for
neutrons and protons in nuclear matter with isospin asymmetry $\alpha =0$
and $0.5$. In Fig. \ref{OPReImEkin}, the resulting energy dependence is
shown for the three nucleon densities $\rho _{\text{B}}=0.08$ fm$^{-3}$
(panel (a)), $0.16$ fm$^{-3}$ (panel (b)), and $0.24$ fm$^{-3}$ (panel (c)).
For all densities, the optical potential shows a strong energy dependence
below $300$ MeV, where it is known that the influences due to ambiguities in
the relativistic form of the NN interaction, the exchange contribution, and
the medium modification due to Pauli blocking are important. The low energy
behavior of the optical potential can in principle be studied in the
generalized relativistic impulse approximation based on the relativistic
meson-exchange model of nuclear force and using the complete set of
Lorentz-invariant NN amplitude \cite%
{horowitz85,murdock87,tjon85,ott88,toki01}. Since many theoretical studies
have shown that data on elastic nucleon-nucleus scattering can be reproduced
by using above optical potential when the nucleon kinetic energy is greater
than about $300$ MeV and that this optical potential also agrees very well
with that extracted from phenomenological analysis of the nucleon-nucleus
scattering data \cite{mcneil83,shepard83,clark83,jin93}, we thus focus in
present work on the higher energy behavior of the isospin-dependent optical
potential. As shown in Fig. \ref{OPReImEkin}, for all three densities
considered here, there is a systematic difference or isospin splitting in
the optical potentials for protons and neutrons in asymmetric nuclear
matter. Specifically, the neutron exhibits a stronger real but weaker
imaginary scalar and vector potentials in neutron-rich nuclear matter.
Furthermore, both the proton and neutron optical potentials become stronger
with increasing density.

\begin{figure}[th]
\includegraphics[scale=0.85]{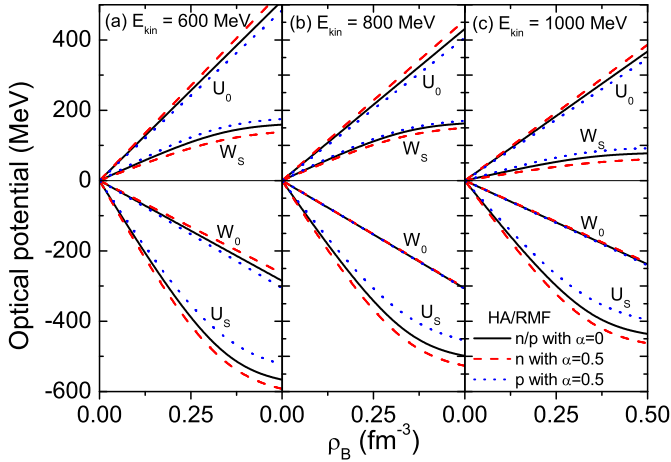}
\caption{{\protect\small (Color online) Density dependence of the real and
imaginary parts of the scalar and vector potentials for neutrons and protons
in nuclear matter with isospin asymmetry }$\protect\alpha =0${\protect\small %
\ and }$0.5${\protect\small \ for the parameter set HA.}}
\label{OPReImDen}
\end{figure}

The density dependence of the real and imaginary parts of the scalar and
vector potentials for neutrons and protons in nuclear matter with isospin
asymmetry $\alpha =0$ and $0.5$ obtained with the parameter set HA is shown
more explicitly in Fig. \ref{OPReImDen} for the three nucleon kinetic
energies of $E_{\mathrm{kin}}=600$ MeV (panel (a)), $800$ MeV (panel (b),
and $1000$ MeV (panel (c). An isospin splitting of the nucleon optical
potential in asymmetric nuclear matter is again clearly seen.

\subsection{Nuclear symmetry potential}

From the Dirac optical potential, a \textquotedblleft Schr\"{o}%
dinger-equivalent potential \textquotedblright\ (SEP) of the following form
is usually introduced \cite{brock78,jaminon80}: 
\begin{equation}
U_{\text{SEP}}=U_{S}^{\mathrm{tot}}+U_{0}^{\mathrm{tot}}+\frac{1}{2M}(U_{S}^{%
\mathrm{tot2}}-U_{0}^{\mathrm{tot2}})+\frac{U_{0}^{\mathrm{tot}}}{M}
\varepsilon,  \label{SEP}
\end{equation}%
where $\varepsilon $ is the nucleon kinetic energy. Using the SEP in the Schr%
\"{o}dinger equation gives the same bound-state energy eigenvalues and
elastic phase shifts as the solution of the upper component of the Dirac
spinor in the Dirac equation using corresponding Dirac optical potential.
The real part of the SEP is then given by 
\begin{equation}
\text{Re}(U_{\text{SEP}})=U_{S}+U_{0}+\frac{1}{2M}%
(U_{S}^{2}-W_{S}^{2}-(U_{0}^{2}-W_{0}^{2}))+\frac{U_{0}}{M}\varepsilon .
\label{ReSEP}
\end{equation}
Above equation corresponds to the nuclear mean-field potential in
non-relativistic models \cite{jaminon89,fuchs05} and allows us to obtain the
following nuclear symmetry potential, i.e., the so-called Lane potential 
\cite{lane62}: 
\begin{equation}
U_{\text{sym}}=\frac{\text{Re}(U_{\text{SEP}})_{n}-\text{Re}(U_{\text{SEP}%
})_{p}}{2\alpha },
\end{equation}%
where Re$(U_{\text{SEP}})_{n}$ and Re$(U_{\text{SEP}})_{p}$ represent,
respectively, the real part of the SEP for the neutron and proton.

\begin{figure}[th]
\includegraphics[scale=1.15]{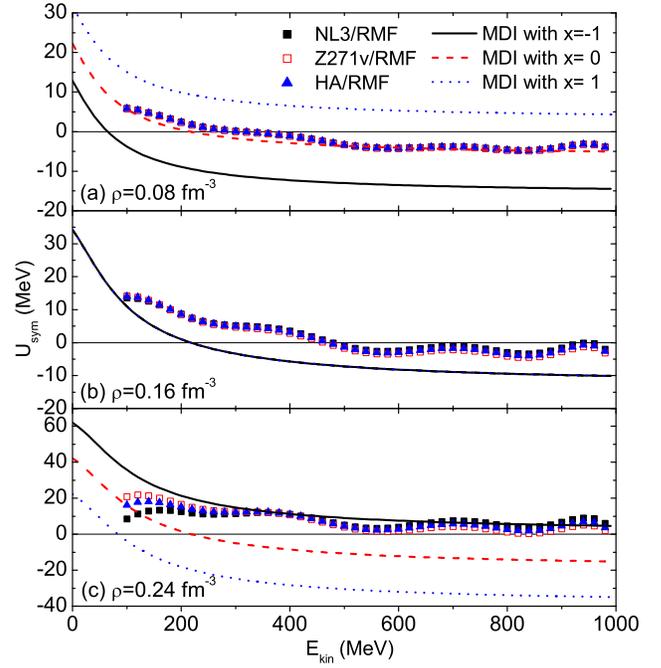}
\caption{{\protect\small (Color online) Energy dependence of the nuclear
symmetry potential using the parameter sets NL3, Z271v, and HA as well as
from the phenomenological interaction MDI with} $x=-1,0,${\protect\small \
and }$1$ at fixed baryon densities of $\protect\rho _{\text{B}}=0.08$ 
{\protect\small \ fm}$^{-3}${\protect\small \ (a), }$0.16${\protect\small \
fm}$^{-3}${\protect\small \ (b), and }$0.24${\protect\small \ fm}$^{-3}$%
{\protect\small \ (c).}}
\label{LaneEkin}
\end{figure}

In Fig. \ref{LaneEkin}, we show the energy dependence of the nuclear
symmetry potential for the parameter sets NL3, Z271v, and HA at fixed baryon
densities of $\rho _{\text{B}}=0.08$ fm$^{-3}$ (panel (a)), $0.16$ fm$^{-3}$
(panel (b)), and $0.24$ fm$^{-3}$ (panel (c)). It is seen that all three
parameter sets give similar nuclear symmetry potential for a nucleon at
kinetic energy higher than about $300$ MeV, i.e., it first decreases with
nucleon kinetic energy and then becomes essentially constant when the
nucleon kinetic energy is above about $500$ MeV. Specifically, the nuclear
symmetry potential starts from about $0$ MeV at lower density of $\rho _{%
\text{B}}=0.08$ fm$^{-3}$ (about half of nuclear saturated density), $4.8$
MeV at normal nuclear matter density ($\rho _{\text{B}}=0.16$ fm$^{-3}$),
and $12$ MeV at higher density of $\rho _{\text{B}}=0.24$ fm$^{-3}$ (about $%
1.5$ time nuclear saturated density) and then saturates to about $-3.8\pm 0.5
$ MeV, $-1.8\pm 1.7$ MeV, and $5.3\pm 3.8$ MeV, respectively, when the
nucleon kinetic energy is greater than about $500$ MeV. The uncertainties in
the saturated values simply reflect the variation in the energy dependence
of the symmetry potential at high energies.

For comparison, we also show in Fig. \ref{LaneEkin} results from the
phenomenological parametrization of the momentum-dependent nuclear
mean-field potential, i.e., MDI interaction with $x=-1$, $0$, and $1$. In
the MDI interaction, the single nucleon potential in asymmetric nuclear
matter with isospin asymmetry $\alpha $ is expressed by \cite%
{das03,li04b,chen04,chen05} 
\begin{eqnarray}
&&U(\rho ,\alpha ,\mathbf{{p},\tau ,{r})}  \notag \\
&=&\left( -95.98-x\frac{2B}{\sigma +1}\right) \frac{\rho _{\tau ^{\prime }}}{%
\rho _{0}}  \notag \\
&&+\left( -120.57+x\frac{2B}{\sigma +1}\right) \frac{\rho _{\tau }}{\rho _{0}%
}  \notag \\
&&+B\left( \frac{\rho }{\rho _{0}}\right) ^{\sigma }(1-x\alpha ^{2})-8\tau x%
\frac{B}{\sigma +1}\frac{\rho ^{\sigma -1}}{\rho _{0}^{\sigma }}\alpha \rho
_{\tau ^{\prime }}  \notag \\
&&+\frac{2C_{\tau ,\tau }}{\rho _{0}}\int d^{3}\mathbf{p}^{\prime }\frac{%
f_{\tau }(\mathbf{{r},{p}^{\prime })}}{1+(\mathbf{p}-\mathbf{p}^{\prime
})^{2}/\Lambda ^{2}}  \notag \\
&&+\frac{2C_{\tau ,\tau ^{\prime }}}{\rho _{0}}\int d^{3}\mathbf{p}^{\prime }%
\frac{f_{\tau ^{\prime }}(\mathbf{{r},{p}^{\prime })}}{1+(\mathbf{p}-\mathbf{%
p}^{\prime })^{2}/\Lambda ^{2}}.  \label{mdi}
\end{eqnarray}%
In the above $\tau =1/2$ ($-1/2$) for neutrons (protons) and $\tau \neq \tau
^{\prime }$; $\sigma =4/3$; and $f_{\tau }(\mathbf{{r},{p})}$ is the phase
space distribution function at coordinate $\mathbf{r}$ and momentum $\mathbf{%
p}$. The parameters $B$, $C_{\tau ,\tau }$, $C_{\tau ,\tau ^{\prime }}$ and $%
\Lambda $ were determined by fitting the momentum-dependence of $U(\rho
,\alpha ,\mathbf{p},\tau ,\mathbf{r})$ to that predicted by the Gogny
Hartree-Fock and/or the Brueckner-Hartree-Fock (BHF) calculations \cite%
{91bomb}, the saturation properties of symmetric nuclear matter and the
symmetry energy of $31.6$ MeV at normal nuclear matter density $\rho
_{0}=0.16$ fm$^{-3}$ \cite{das03}. The incompressibility $K_{0}$ of
symmetric nuclear matter at $\rho _{0}$ is set to be $211$ MeV. The
different $x$ values in the MDI interaction are introduced to vary the
density dependence of the nuclear symmetry energy while keeping other
properties of the nuclear EOS fixed \cite{chen05}. We note that the energy
dependence of the symmetry potential from the MDI interaction is consistent
with the empirical Lane potential at normal nuclear matter density and low
nucleon energies \cite{li04a} and has been used in the transport model for
studying isospin effects in intermediate-energy heavy ion collisions induced
by neutron-rich nuclei \cite{li04b,chen04,chen05}.

It is seen from Fig. \ref{LaneEkin} that results from RIA at lower density
of $\rho =0.08$ fm$^{-3}$ are comparable to those from the MDI interaction
with $x=0$, while at higher baryon density of $\rho _{\text{B}}=0.24$ fm$%
^{-3}$ they are comparable to those from the MDI interaction with $x=-1$. At
normal nuclear matter density, the MDI interaction, which gives same results
for different $x$ values by construction, is seen to lead to smaller nuclear
symmetry potential at high nucleon kinetic energies compared with present
results from the RIA based on empirical NN scattering amplitude and the
nuclear scalar density from the relativistic mean-field theory. We note that
our results agree surprisingly well with those of DBHF by Fuchs \textsl{et
al.} \cite{fuchs04,fuchs}.

\begin{figure}[th]
\includegraphics[scale=0.95]{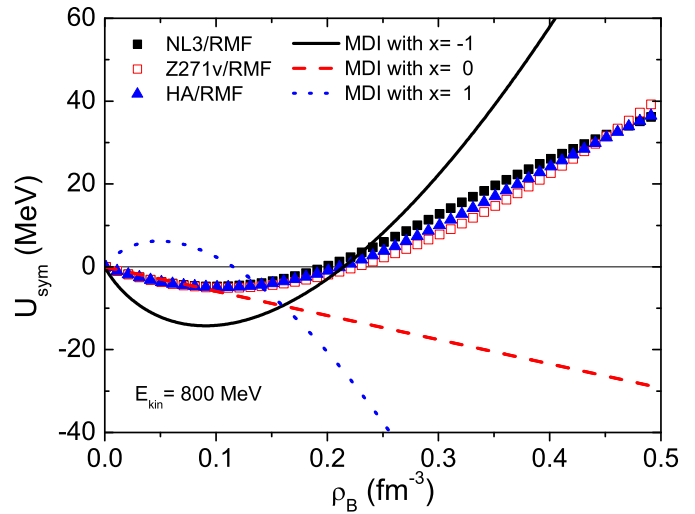}
\caption{{\protect\small (Color online) Density dependence of the nuclear
symmetry potential using the parameter sets NL3, Z271v, and HA as well as
from the MDI interaction with }$x=-1${\protect\small , }$0${\protect\small ,
and }$1$ at a fixed nucleon kinetic energy of 800 MeV.}
\label{LaneDen}
\end{figure}

For the density dependence of the nuclear symmetry potential using the
parameter sets NL3, Z271v, and HA at a fixed high nucleon kinetic energy of $%
800$ MeV, it is shown in Fig. \ref{LaneDen} together with corresponding
results from the MDI interaction with $x=-1$, $0$, and $1$. It is clearly
seen that the nuclear symmetry potential from all parameter sets NL3, Z271v,
and HA changes from negative to positive values at a fixed baryon density of
about $\rho _{\text{B}}=0.22$ fm$^{-3}$ and then increases almost linearly
with baryon density. Furthermore, the nuclear symmetry potential depends not
much on the choice of the parameter sets NL3, Z271v, and HA. At such high
nucleon kinetic energy, the nuclear symmetry potential from the MDI
interaction with $x=0$ reproduces nicely the results from the RIA when $\rho
_{\text{B}}\lesssim 0.1$ fm$^{-3}$ as in its energy dependence at low
densities shown in Fig. \ref{LaneEkin}. The two differ strongly, however, at
high densities. The MDI interaction with both $x=-1$ and $1$, on the other
hand, show very different density dependence from present RIA results.

\section{summary}

\label{summary}

We have evaluated the Dirac optical potential for neutrons and protons in
asymmetric nuclear matter based on the relativistic impulse approximation
with empirical NN scattering amplitude and the scalar and vector densities
from the relativistic mean-field theory. We find that the nuclear symmetry
potential derived from the resulting Schr\"{o}dinger-equivalent potential is
not very sensitive to the parameter sets used in the relativistic mean-field
calculation, particularly at low densities and high nucleon energies,
although the latter give very different nuclear scalar densities at high
baryon densities in both symmetric and asymmetric nuclear matters.
Furthermore, the nuclear symmetry potential at a fixed density becomes
almost constant when the nucleon kinetic energy is greater than about $500$
MeV. For such high energy nucleon, our study shows that the density
dependence of its nuclear symmetry potential is weakly attractive at low
densities but become increasingly repulsive as nuclear density increases.
Results presented in present study thus provide important constraints on the
high energy behavior of the nuclear symmetry potential in asymmetric nuclear
matter, which is an important input to the isospin-dependent transport model 
\cite{li04b,baran05} in studying heavy-ion collisions induced by radioactive
nuclei at intermediate and high energies. They are also useful in future
studies that extend the Lorentz-covariant transport model \cite{ko87,mosel92}
to include explicitly the isospin degrees of freedom.

\begin{acknowledgments}
This work was supported in part by the National Natural Science Foundation
of China under Grant Nos. 10105008 and 10575071 (LWC), the US National
Science Foundation under Grant No. PHY-0457265 and the Welch Foundation
under Grant No. A-1358 (CMK), and the US National Science Foundation under
Grant Nos. PHY-0354572 and PHY-0456890, and the NASA-Arkansas Space Grants
Consortium Award ASU15154 (BAL).
\end{acknowledgments}

\end{document}